\documentclass[aip,twocolumn,10pt,letterpaper,bibnotes,floatfix,raggedbottom]{revtex4-1}
\usepackage{subfigure}
\usepackage{dcolumn}
\usepackage{bm}
\usepackage{graphicx}
\usepackage{sidecap}
\usepackage{amsmath}
\usepackage{amssymb}
\usepackage{physics}
\usepackage[protrusion=true,expansion]{microtype}
\usepackage{times}
\usepackage{multirow}
\usepackage{floatrow}
\usepackage{float}
\floatstyle{plaintop}
\restylefloat{table}
\usepackage{dsfont} 

\newcommand*{\GtrSim}{\smallrel\gtrsim}

\newcommand*{\LtrSim}{\smallrel\lesssim}

\makeatletter
\newcommand*{\smallrel}[2][.8]{%
  \mathrel{\mathpalette{\smallrel@{#1}}{#2}}%
}
\newcommand*{\smallrel@}[3]{%
  \sbox0{$#2\vcenter{}$}%
  \dimen@=\ht0 %
  \raise\dimen@\hbox{%
    \scalebox{#1}{%
      \raise-\dimen@\hbox{$#2#3\m@th$}%
    }%
  }%
}
\makeatother

\begin{document}

\preprint{This article may be downloaded for personal use only. Any other use requires prior permission of the author and AIP Publishing. }

\preprint{This article appeared in Appl. Phys. Lett. \textbf{113}, 092901 (2018) and may be found at \href{https://aip.scitation.org/doi/10.1063/1.5046080}{https://aip.scitation.org/doi/10.1063/1.5046080}.}

\vspace*{20pt}

\title{Electromechanical control of polarization vortex ordering in an interacting ferroelectric-dielectric composite dimer}

\author{John Mangeri}
\email{mangeri@fzu.cz}
\affiliation{\hbox{Institute of Physics, Czech Academy of Sciences, Na  Slovance  2,  182  21  Prague  8, Prague,  Czech  Republic}}
\affiliation{\hbox{Department of Physics, University of Connecticut, Storrs, Connecticut 06269, USA}}

\author{S. Pamir Alpay}
\affiliation{\hbox{Department of Physics, University of Connecticut, Storrs, Connecticut 06269, USA}}
\affiliation{Department of Materials Science \& Engineering, and Institute of Materials Science, University of Connecticut, Storrs, Connecticut 06269, USA}

\author{Serge Nakhmanson}
\affiliation{\hbox{Department of Physics, University of Connecticut, Storrs, Connecticut 06269, USA}}
\affiliation{Department of Materials Science \& Engineering, and Institute of Materials Science, University of Connecticut, Storrs, Connecticut 06269, USA}

\author{Olle G. Heinonen}
 \affiliation{\hbox{Material Science Division, Argonne National Laboratory, Lemont, Illinois 60439, USA}}
 \affiliation{Center for Hierarchical Material Design, Northwestern-Argonne Institute of Science and Engineering, Northwestern University, Evanston, Illinois 60201, USA}

\date{June 25, 2018}

\begin{abstract}
Using a free-energy based computational model, we have investigated the response of a system comprised of two interacting ferroelectric 
nanospheres, embedded in a dielectric medium, to a static external electric field.
The system response is hysteretic and tunable by changing the inter-particle distance and the orientation of the applied field,
which strongly modulates the field-driven long-range elastic interactions between the particles that propagate through the dielectric matrix.
At small separations, the sensitivity of the system behavior with respect to the electric field direction originates
from drastically different 
configurations of the local vortex-like polarization states in the ferroelectric particles.
This suggests new routes for the design of composite metamaterials whose dielectric properties can be controlled 
and tuned
by selecting the mutual arrangement of their ferroelectric components.

\end{abstract}

\maketitle



%
Ferroelectric-dielectric composite systems display a wide range of functional behavior.
For particulate dispersed phases, precise control of the inter-particle spacing and arrangement \cite{Feng2017}, 
particle shape morphology \cite{Hao2014, Caruntu2015}, surface crystallography \cite{Wang2018} and concentration \cite{Hao2017}, as well as 
selection of materials for both phases (with the matrix being of ceramic\cite{Airiioaei2017, Nenasheva2017} 
or polymeric origin\cite{Huang2010, Paniagua2014, Hao2017, Feng2017}), can lead to designs with preprogrammed
properties and novel functionalities.
In particular, remarkable behavioral features, such as superparaelectricity \cite{Huang2010}, enhanced dielectric tunability \cite{Zhou2008} and 
energy storage density \cite{Huang2010, Zhang2015, Liu2017}, multi-state ferroelectric (FE) switching \cite{Gruverman2008, Li2017}, and reduced conductivity and 
loss \cite{Airiioaei2017, Nenasheva2017} were already demonstrated in ferroelectric-dielectric composites, which makes these systems attractive for 
numerous future technological applications \cite{Alitalo2009}.
%

\begin{figure}[b!]
\centering
\includegraphics[width=0.94\linewidth]{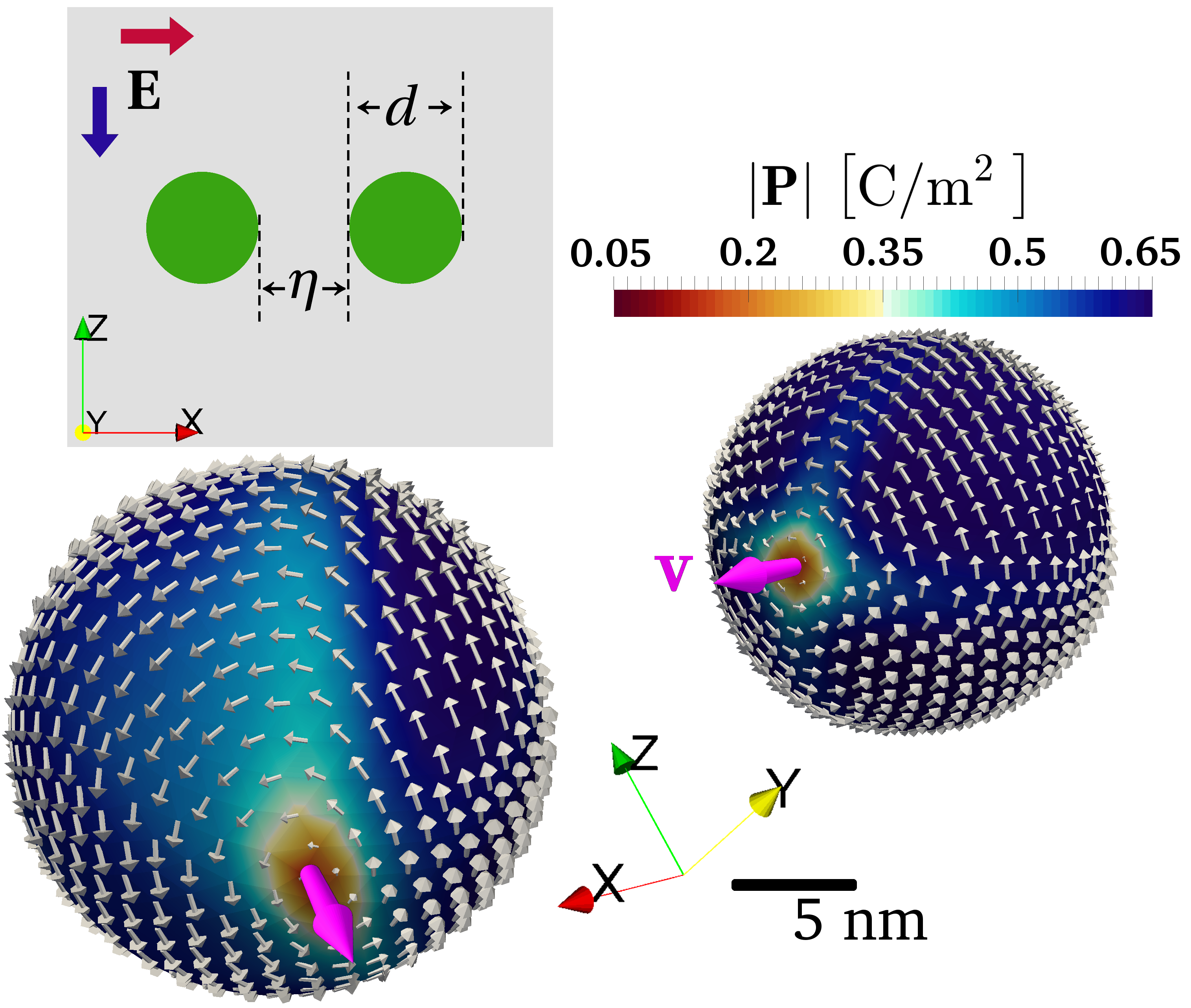}
\caption{
Top panel: diagram (not to scale) of the interacting PTO particle dimer of diameter $d$ separated by a surface-to-surface distance $\eta$, embedded in a 
matrix of a linear elastic dielectric (here STO). 
%
%
The electric field directions are shown as blue ($\hat{\mathbf{E}} || \hat{\mathbf{z}}$) or red ($\hat{\mathbf{E}} || \hat{\mathbf{x}}$) arrows, with the latter 
being parallel to the axis connecting the particles. 
The bottom panel shows a typical weakly-correlated polarization vortex-like state of the dimer with $\eta = 10$ nm after an energy minimization in zero applied field. 
Director $\mathbf{v}$ (pink arrow) is an axial vector representing the orientation of the cylindrical polar-vortex core.
}\label{fig:dimer_diagram_1}
\end{figure}

%
A number of synthetic techniques have been recently used to fabricate nanoparticles of 
binary\cite{Jana2004,Jun2006,Tang2016} and ternary\cite{Park2005,Tyson2014,Papaefthymiou2015,Caruntu2015,Zhang2017} 
oxides in a variety of shapes and sizes.
The properties of these particles, as well as the nature of FE ordering in them, have been characterized with multiple 
experimental approaches, including Bragg coherent diffraction imaging\cite{Karpov2017} and second harmonic generation 
microscopy\cite{Hertel2014,Ma2017,Timpu2017} for sizes of above 50--60 nm, and aberration corrected transmission
electron microscopy (TEM) \cite{Polking2012,Yadav2016} and local piezoforce microscopy 
for particles of
to 5--15 nm in size \cite{Rodriguez2009,Polking2012}.
Both monodomain \cite{Polking2012} and vortex-like \cite{Rodriguez2009,Karpov2017,Li2017} polarization topologies were observed 
in the FE particles, with the latter patterns expected to be utilized in memory elements for new paradigms in computing \cite{Scott2007, Gruverman2008}.
Formation of topological polarization patterns in \emph{isolated} FE particles has also been studied with different coarse-grained theoretical approaches, 
such as phase-field \cite{Wu2014, Liu2017, Mangeri2017, Yadav2016}, semi-analytic \cite{Akdogan2007a, Glinchuk2008, Martelli2015, Qiu2015}, 
and effective hamiltonian \cite{Naumov2004, Naumov2008} methods.
For example, in previous work \cite{Mangeri2017,Pitike2018}, we evaluated size and shape dependent phase transitions in \emph{individual} FE 
nanoparticles embedded in a dielectric matrix, observing a monodomain to vortex-like to polydomain
sequence of transformations for the increasing particle size.
However, in a composite system, this situation corresponds to a highly dispersed arrangement of particles within the matrix, with negligible 
electrostatic and elastic interactions between them.
On the other hand, it is expected that more promising functional behavior can be obtained in complex systems consisting of many 
\emph{interacting} particles, 
the aggregate responses of which should be heavily influenced by the particle size, shape, anisotropy and
mutual arrangement \cite{Feng2017,Cai2017}.
As accurate treatment of composite materials including numerous interacting FE particles presents a formidable computational challenge even on the
coarse-grained level of theory, the next step towards developing a better understanding of such compounds involves investigation of
systems with few interacting particles to contrast their behavior against the single-particle case.
Here, using the time-dependent Landau-Ginzburg-Devonshire (TDLGD) approach \cite{Li2001, Mangeri2017}, 
we explore a dimer system, consisting of two PbTiO$_3$ (PTO) nanospheres embedded in a linear dielectric medium represented 
by SrTiO$_3$ (STO). 
Our main goals include determination of a zero-field polarization pattern within the dimer as a function of the inter-particle separation $\eta$, followed by 
evaluation of the quasi-static response of the dimer system 
to an electric field that can be applied in different directions (as shown in the top panel of Fig.~\ref{fig:dimer_diagram_1}).
Throughout this study, both particle diameters are kept fixed at $d = 10$ nm: the size identified in our previous investigation\cite{Mangeri2017} 
as the one that supports a zero electric field vortex-like polarization state.
Following the TDLGD approach, the Cartesian components $k=1,2,3$ of the local polarization field, $\mathbf{P}(\mathbf{r})$, are evolved in time as 
\begin{equation}\label{TDLGD}
\frac{\partial P_k}{\partial t} =  - \Gamma \frac{\delta {\mathcal F}}{\delta P_k},
\end{equation}
where ${\mathcal F}$ is the total electric enthalpy of the composite, and $\Gamma$ is a coefficient related to the FE domain wall mobility \cite{Hlinka2007}. 
Detailed expressions for all of the energy terms included in ${\mathcal F}$ are provided in the Supplemental Materials (SM), together with the utilized materials parameters
for PTO and STO.
We assume that during the system time evolution, elastic strains relax at much faster time-scales than $\mathbf{P}$ \cite{Li2001, Li2002395}.
Therefore, at each time step, for fixed polarization, 
the strain fields $\varepsilon_{ij} = \frac{1}{2} \left(\partial u_i/\partial x_j + \partial u_j/\partial x_i \right)$ 
can be obtained from the mechanical equilibrium equation
\begin{equation}\label{stress-div}
\frac{\partial}{\partial x_j} C_{ijkl} \left( \varepsilon_{kl} - Q_{klmn} P_m P_n \right) = 0,
\end{equation}
where $C_{ijkl}$ and $Q_{ijkl}$ are the elastic stiffness and electrostrictive tensor components, respectively, 
and $\mathbf{u}(\mathbf{r})$ is the 
elastic displacement field.
The 
electrostatic potential $\Phi(\mathbf{r})$ is computed from the Poisson equation
\begin{equation}\label{poisson}
\frac{\partial}{\partial x_j}\left(\epsilon_b \frac{\partial \Phi}{\partial x_j} \right) = \frac{\partial P_k}{\partial x_k},
\end{equation}
where $\epsilon_b$ is the background dielectric tensor \cite{Hlinka2006}, 
assumed to be isotropic. 
%

%
%
We solve Eqs.~(\ref{TDLGD}--\ref{poisson}) using finite-element methods within   
the open-source MOOSE framework\cite{Gaston2009} and the \textsc{Ferret} module \cite{FerretLink} 
developed by the authors. 
%
%

In the absence of an applied external electric field, the boundary conditions on the six sides of the system volume (chosen to be sufficiently large with the dimer at its center) are $\mathbf{u}=0$, and  $\Phi= 0$. 
%
In the presence of an applied $\mathbf{E}$ field, a potential difference $\Delta \Phi = \Phi_1 - \Phi_2$ 
is applied to two opposing sides of the box.
The initial condition on $\mathbf{P}(\mathbf{r})$ is chosen as a random field of small local polarizations such that $\langle \mathbf{P} \rangle \approx 0$, i.e., mimicking 
the high-temperature paralectric state. 
As ${\mathbf {P}}$ is evolved in time [Eq.~(\ref{TDLGD})], 
energy is dissipated from the system, with the simulation terminated when the relative change 
in ${\mathcal F}$ between consecutive time steps is approximately $0.01\%$.
%
%

\begin{figure*}[htpb!]
\centering
\includegraphics[width=0.8\linewidth]{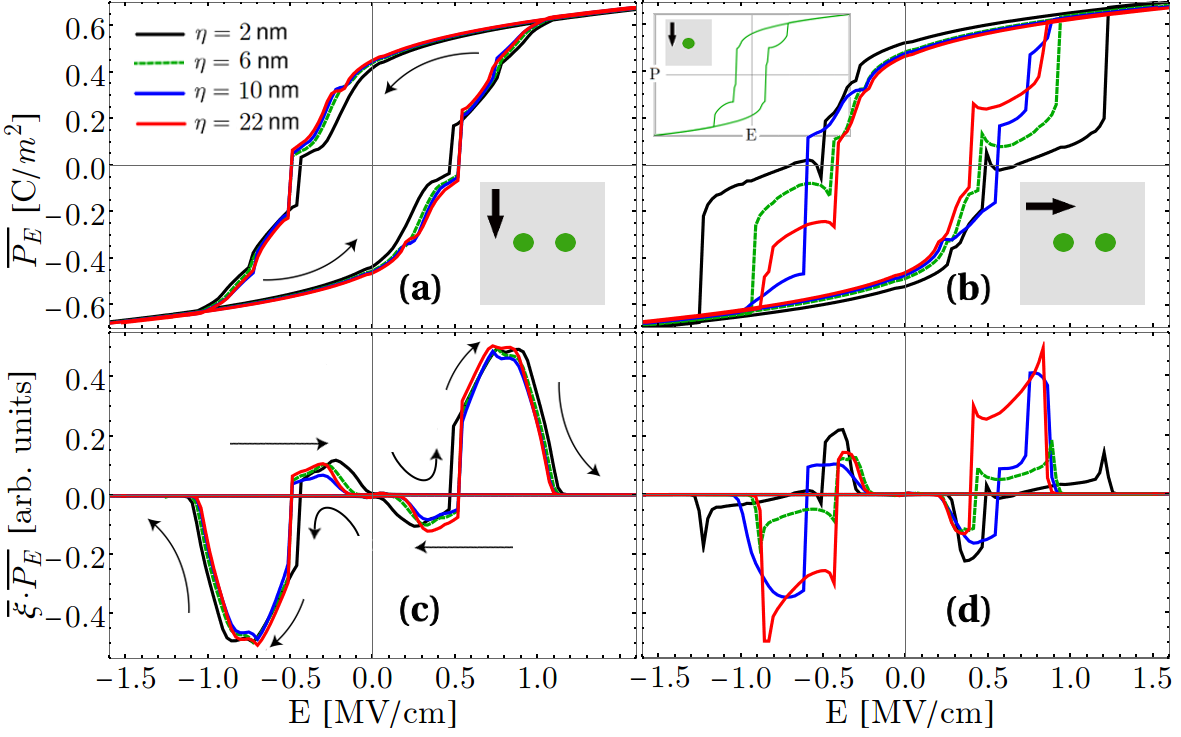}
\caption{
$P$-$E$ loops for the averaged polarization component along the direction of the applied electric field $P_E$ for (a): $\hat{\mathbf{E}} || \hat{\mathbf{z}}$ 
and (b): $\hat{\mathbf{E}} || \hat{\mathbf{x}}$, and $\eta = 2$~nm (black), $6$ ~nm (green), $10$~nm (blue), and $22$~nm (red). 
Panels (c) and (d) depict the magnitude of the averaged winding number density $\xi$ weighted with the average polarization along the field $P_E$
as a function of the electric field for the configurations (a) and (b), respectively.
The upper inset in (b) shows the $P$-$E$ loop of a single sphere with $d = 10$~nm calculated with the same approach (from Ref.~\onlinecite{Mangeri2017}).
}\label{fig:hyst}
\end{figure*}

%
An example of a dimer equilibrium state for $\eta = 10$ nm under the $\mathbf{u}, \Phi \to 0$ boundary conditions is presented in Fig.~\ref{fig:dimer_diagram_1}.
The polarization field aligns tangentially to the particle surface to minimize 
surface charges, 
creating a vortex-like pattern in each of the two spheres.
Such an equilibrium structure is different from the flux-closure points typically observed at intersections of four $90^\circ$ 
FE domain walls\cite{McQuaid2014, Wu2014}. 
The pattern of Fig.~\ref{fig:dimer_diagram_1} has diffuse domain walls, with the $\mathbf{P}(\mathbf{r})$ distribution continuously rotating 
about a central axis \cite{Mangeri2017, Karpov2017} represented by an \emph{axial} director $\mathbf{v}$.
%
The cylindrical vortex-core region 
penetrates through the sphere and usually has polarization magnitudes that are lower than at the surfaces. 
%
This behavior is different from that of ferromagnetic vortices, 
in which the magnetization density magnitude is constant\cite{Lee2014,Zhou2015} well below $T_{\rm C}$.
For inter-particle spacings $\eta \LtrSim 10$ nm, the core-region orientations in the two particles can become strongly correlated.
Specifically, the angles between the directors $\mathbf{v}$ of the two spheres can adopt a discrete set of values, 
suggesting that certain mutual core-region orientations within the dimer are more energetically preferred than others.
In contrast with the single particle case, in some instances we also observe the development of large polarization, comparable to the surface values,
in the vortex-core regions of both spheres. 
%
%
Some examples of such polarization patterns are shown in the SM for $\eta = 2$ nm. 
%
%
%
%
For large inter-particle distances $\eta \GtrSim 10$ nm, we observe no pronounced correlations between the core-region orientations
of the dimer particles; 
%
the orientation of the director $\textbf{v}$ in each sphere depends mostly on the specific configuration of $\mathbf{P}(\mathbf{r})$ generated by 
the random initial condition.
Therefore, in the absence of an externally applied $\mathbf{E}$, the dimer particles do not strongly interact with each other: 
the stray fringing electric fields from the surface and volume charge distributions of each sphere are effectively screened by 
the high dielectric permittivity ($\epsilon_b = 300\epsilon_0$) of the surrounding matrix, and the strain fields propagating from 
the surface of each sphere into the matrix decay within a short distance.

We next turn to the analysis of the quasi-static dimer $P$-$E$ curves \cite{Mangeri2017,Boardman2004}. 
In the previous study, the $P$-$E$ loops of an isolated PTO sphere with $d = 10$ nm embedded in STO were found to exhibit shelves of 
roughly constant $\overline{P_E} = \overline{\textbf{P} \cdot \hat{\textbf{E}}}$ at smaller applied fields ($\overline{X}$ denotes volume-average of a scalar field $X$), 
which was associated with the presence of a vortex-like phase that transformed into a uniform
monodomain phase when the field was increased, as shown in the insert of Fig.~\ref{fig:hyst}(b).
A useful measure of the system vorticity is the FE-volume-averaged 
winding number density $\xi = |\textbf{P} \cdot \left(\nabla \times \textbf{P}\right)|$; $\xi\approx0$ 
%
in the monodomain phase, and saturates to a constant value in the polydomain phase,
however, in the adjoining vortex-like phase region it goes through a sharp maximum \cite{Mangeri2017}. 
The $P$-$E$ loops presented in Fig.~\ref{fig:hyst} (a) and (b) are computed for 
$\eta = 2, 6, 10$, and $22$~nm
for both orientations of the applied electric field shown in the top panel of Fig.~\ref{fig:dimer_diagram_1}.
For each point of the loop, the components of local $P_E(\mathbf{r})$ are averaged over both spheres. 
The averaged winding number density $\bar{\xi}$ is presented in panels (c) and (d) of Fig.~\ref{fig:hyst} weighted with the average polarization along the field $\overline{P_E}$ as a function of the applied field magnitude $E$.
%
%

%
The orientation of the applied field relative to the dimer axis profoundly influences the shape of the $P$-$E$ response. 
With the field perpendicular to the axis connecting the particles ($\hat{\mathbf{E}} || \hat{\mathbf{z}}$) [Fig.~ref{fig:hyst}(a) and (c)], 
both the $\overline{P_E}$ and $\bar{\xi} \cdot \overline{P_E}$ dependencies exhibit only minor changes with increasing inter-particle spacing $\eta$.
In particular, the transition between monodomain and vortex-like states occurs at the same value of $E \approx$ 0.2--0.3~MV/cm for all values of $\eta$. 
As the applied field switches sign, a small domain of correlated $\mathbf{P}$ along the direction of $\mathbf{E}$ 
begins to nucleate within the vortex-like phase in each sphere.
With the increasing strength of the applied field, these domains grow until reaching saturation, which results in a gradual
decrease and then complete disappearance of $\bar{\xi}$.
This behavior seems to be universal and not strongly dependent on the value of inter-particle spacing $\eta$. 

\begin{figure*}[htpb!]
\centering
\includegraphics[width=0.90\linewidth]{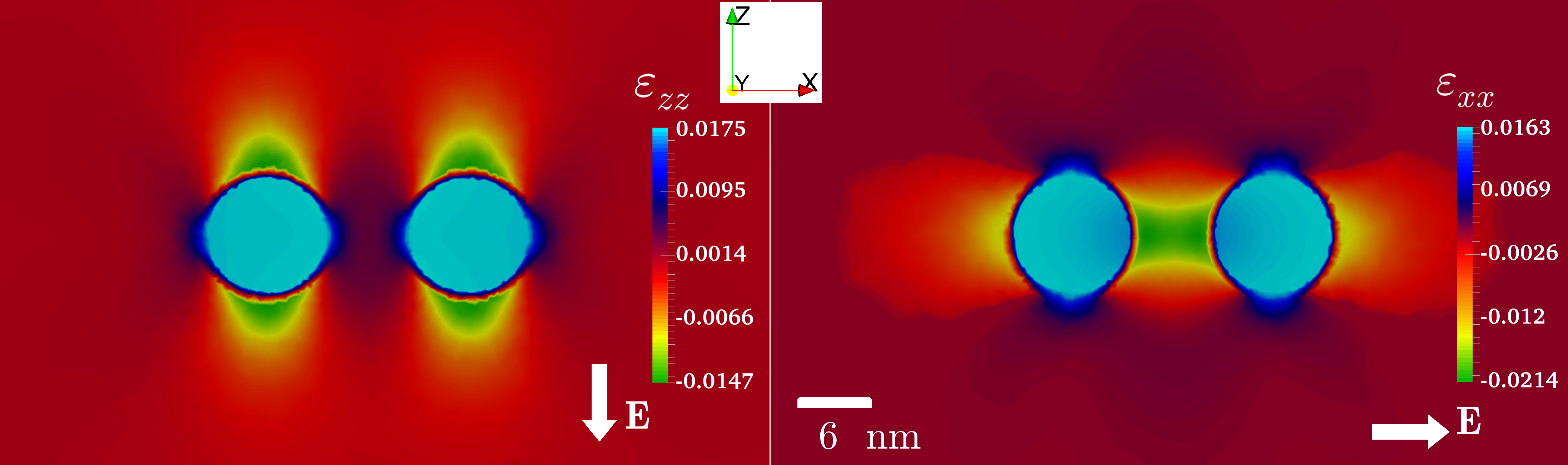}
\caption{
Color maps of the diagonal components of the $\varepsilon_{ii}$ in the $x$-$z$ plane in the 
dimer separated by $\eta = 6$ nm, with the field $E$ = 1.0~MV/cm applied along $-\hat{\mathbf{z}}$ (left panel) and $-\hat{\mathbf{x}}$ (right panel). 
The tensile strain is nearly constant inside the spheres, which is consistent with a constant-polarization state along the applied electric field direction. 
As the spheres expand slightly into the matrix, it develops compressed regions that are disconnected when $\hat{\mathbf{E}} || \hat{\mathbf{z}}$ (left panel) 
and joined when $\hat{\mathbf{E}} || \hat{\mathbf{x}}$ (right panel).
Note the different color bar scales in the panels.
}\label{fig:strain_anisotropy}
\end{figure*}

\begin{figure}[b!]
\centering
\includegraphics[width=0.62\linewidth]{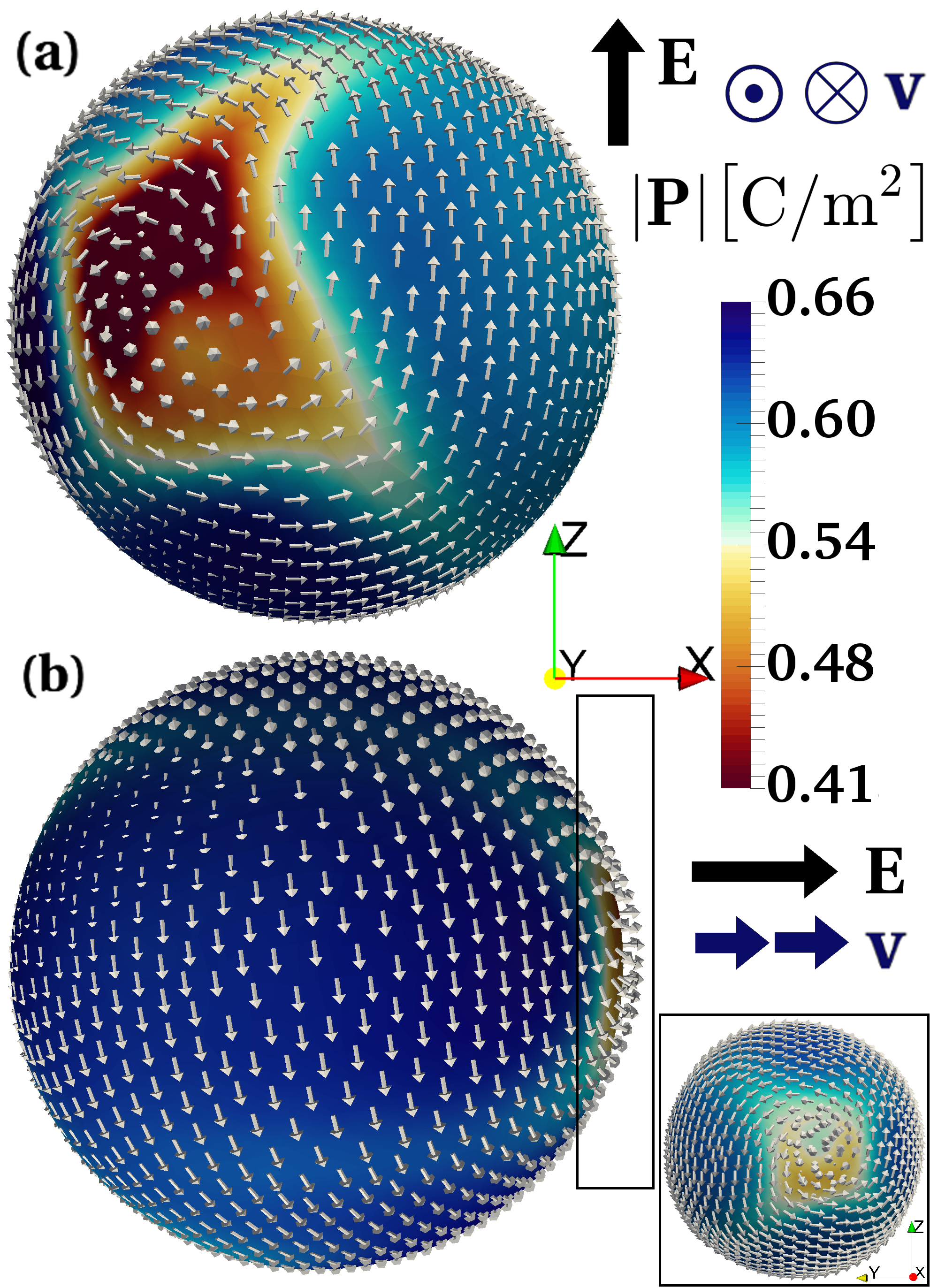}
\caption{
$\mathbf{P}(\mathbf{r})$ patterns in one of the dimer particles for $\eta = 6$ nm, $|E| = 0.7$ MV/cm and (a) $\hat{\mathbf{E}} || \hat{\mathbf{z}}$, 
or (b) $\hat{\mathbf{E}} || \hat{\mathbf{x}}$. 
The depressed magnitude of the core-region polarization is highlighted by the color map.
The inset to panel (b) shows the vortex-like $\mathbf{P}(\mathbf{r})$ pattern with $\mathbf{v} || \mathbf{E}$.
}\label{fig:vortex_ordering_control}
\end{figure}

When the field is applied along the axis connecting the particles ($\hat{\mathbf{E}} || \hat{\mathbf{x}}$), as shown in panels (b) and (d) of Fig.~\ref{fig:hyst}, 
a radically different system behavior is observed, as manifested by the appearance of shelf-like features in both $\overline{P_E}(E)$ 
and $\bar{\xi}(E)$ curves, indicating the presence of the vortex-like phase.
The transition into the vortex-like state happens at the same value of $E$ as in the previous case, i.e., 0.2--0.3~MV/cm. 
The width of the shelf-like feature on the $\overline{P_E}(E)$ curve is roughly inversely proportional to the value of $\eta$,
as can be seen from the comparison of the loops for $\eta = 2$~nm and $\eta = 6$~nm [see Fig.~\ref{fig:hyst} (b)].
The $\bar{\xi}(E)$ dependence exhibits sharp peaks on both ends of the shelf that are especially
pronounced for $\eta = 2$~nm and $\eta = 6$~nm curves and can serve as good indicators for an onset of a phase transition.
Inspection of the corresponding $\mathbf{P}(\mathbf{r})$ distributions shows that these peaks arise during a formation of \emph{multiple} vortex-like patterns in each sphere. 
These patterns then merge into a single vortex with a core direction aligned with the direction of the field; some images illustrating the process are presented in the SM.
For $\eta \GtrSim 20$ nm, the system $P$-$E$ loops become similar to that of an isolated sphere of identical size,
cf.\ the $\eta = 22$~nm curve and the curve in the inset of Fig.~\ref{fig:hyst}(b).
In order to better understand the origins of the varying dimer system response under different orientations of the applied field
we examine the behavior of elastic strain fields as a function of the direction of $\mathbf{E}$.
In Fig.~\ref{fig:strain_anisotropy}, color maps of the diagonal components of the elastic strain tensor $\varepsilon_{ii}$ along the field are presented for the case 
of inter-particle separation of 6 nm and the field magnitude of 1.0~MV/cm applied in the $-\hat{\mathbf{z}}$ direction (left panel) and $-\hat{\mathbf{x}}$ direction (right panel).
At this electric field strength, both particles in each configuration are in the monodomain state, with the polarization parallel to the applied field. 
Due to the nearly constant polarization, the elastic strain inside each sphere along the field direction $i$ is tensile, $\varepsilon_{ii} \approx 0.0165\%$. 
Therefore, the matrix region \emph{in between} the particles experiences tensile strain for $\hat{\mathbf{E}} || \hat{\mathbf{z}}$ and compressive
strain for $\hat{\mathbf{E}} || \hat{\mathbf{x}}$, as shown, respectively, in the left and right panels of Fig.~\ref{fig:strain_anisotropy}.
Elastic distortions extend for $\sim$10~nm from the sphere surfaces into the matrix and decay to zero at the boundaries of the computational domain (see SM for an example).
The strain transfer mechanism in the composite arises from the electromechanical coupling between the FE particles and the dielectric medium: the applied electric field elongates 
the particles slightly along the direction of the field, which in turn causes elastic distortions within the matrix.
For the $\hat{\mathbf{E}} || \hat{\mathbf{x}}$ case, the strain in the region between the spheres is approximately additive
for small separations $\eta$, which leads to a strong elastic coupling between the spheres. 
In turn, this results in a substantial decrease of $\overline{P_E}$, compared to the $\hat{\mathbf{E}} || \hat{\mathbf{z}}$ case:
e.g., $\overline{P_E} \equiv \overline{P_z} = 0.35$ vs.\ $\overline{P_E} \equiv \overline{P_x} = 0.20$ C/$\mathrm{m}^2$ for $\eta = 6$~nm and $E = 0.7$ MV/cm,
as shown in panels (a) and (b), respectively, of Fig.~\ref{fig:hyst}.
The origin of this difference stems from contrasting orientations of the particle vortex cores with respect to $\mathbf{E}$, as shown in Fig.~\ref{fig:vortex_ordering_control}.
In the $\hat{\mathbf{E}} || \hat{\mathbf{z}}$ case, shown in panel (a), core orientation vectors $\mathbf{v}$ of both spheres are perpendicular to $\mathbf{E}$
and it is the large local polarization from the surface regions that contributes to the average value of $P_E$.
In contrast, in the $\hat{\mathbf{E}} || \hat{\mathbf{x}}$ case, shown in panel (b), the directors $\mathbf{v}$ of both spheres are aligned along $\textbf{E}$,
with only the core-region polarization, which is greatly suppressed compared to the surface regions 
contributing to $\overline{P_E}$.
Therefore, it is the $\mathbf{v} || \hat{\mathbf{E}} || \hat{\mathbf{x}}$ system orientation that gives rise to the shelf-like features 
with depressed near-constant $\overline{P_E}$ in Fig.~\ref{fig:hyst}(b) at small inter-particle separations.
As the electric field increases, the core-region polarization saturates and then proliferates in the $y$-$z$ plane until a monodomain state is formed and 
vorticity vanishes.

%
We have investigated the influence dimer size and electric field orientation on the ordering and response of a FE-dielectric composite system. 
%
Our work demonstrates that the dimer size and electric field direction can be used to control the dielectric response in such composite systems, with potential applications in electronic and elecctromechanical devices. 
%

%
\section*{Supplemental Material}
See supplemental material for additional details on the calculations and supporting information.
\section*{Acknowledgments}
JM acknowledges funding under the Ministry of Education, Youth and Sports of the Czech Republic, grant number CZ.02.2.69/0.0/0.0/16$\_$027/0008215.
OH was funded by the Department of Energy, Office of Science Basic Energy Sciences Division of Materials Science and Engineering.
%
%
JM is grateful to A.\ Jokisaari for assistance with calculations.
%

\bibliographystyle{apsrev4-1}
\bibliography{references} 

\begin{thebibliography}{50}
\bibitem [{Fer()}]{FerretLink}%
  \BibitemOpen
  \href@noop {} {}\bibinfo {note} {Ferret is available at
  bitbucket.org/mesoscience/ferret.}\BibitemShut {Stop}%
\end{thebibliography}

\end{document}